# Non-congruence of Thermally Driven Structural and Electronic Transitions in $VO_2$

Joyeeta Nag, Richard F. Haglund Jr.

*Department of Physics and Astronomy*

*Vanderbilt Institute of Nanoscale Science and Engineering,*

*Vanderbilt University, Nashville, TN 37235*

E. Andrew Payzant[1], Karren L. More[2]

[1]*Center for Nanophase Materials Sciences and* [2]*SHaRE Program*

*Oak Ridge National Laboratory*

*Oak Ridge, TN  37831*

**Coupled structural and electronic phase transitions underlie the multifunctional properties of strongly-correlated materials. For example, colossal magnetoresistance[1,2] in manganites involves phase transition from paramagnetic insulator to ferromagnetic metal linked to a structural Jahn-Teller distortion[3]. Vanadium dioxide ($VO_2$) likewise exhibits an insulator-to-metal transition (IMT) at ~67°C with abrupt changes in transport and optical properties and coupled to a structural phase transition (SPT) from monoclinic to tetragonal[4]. The IMT and SPT hystereses are signatures of first-order phase transition tracking the nucleation to stabilization of a new phase. Here we have for the first time measured independently the IMT and SPT hystereses in epitaxial $VO_2$ films, and shown that the hystereses are not congruent. From the measured volume fractions of the two phases in the region of strong correlation, we have computed the evolving dielectric function under an effective-medium approximation. But the computed dielectric functions could not reproduce the measured IMT, implying that there is a strongly correlated metallic phase that is not in the stable rutile structure, consistent with Qazilbash et al[5]. Search for a corresponding macroscopic structural intermediate also yielded negative result.**

The complex physics of $VO_2$ phase transition has long been debated[6-10]. Unlike other strongly correlated materials[11] exhibiting IMT, such as $V_2O_3$[12,13], where phase transitions are satisfactorily explained by the Mott mechanism alone, the IMT in $VO_2$ is complicated by accompanying spin-Peierls instability[14] that leads via strong electron-phonon coupling to the formation of spin-singlet states via the bonding-antibonding splitting of the $d_{x^2-y^2}$ bands.

A critical question so far unanswered is whether the structural and electronic phase transitions in VO$_2$ occur congruently. For non-equilibrium phase transformations induced by ultrashort laser pulses, the electronic IMT leads the SPT [15], and the SPT presents a kinetic bottleneck for the transition [16]. For the shortest attainable excitation pulses, a coherent phonon associated with a breathing mode of the VO$_2$ lattice appears simultaneously with the IMT [17]. Ultrafast electron diffraction measurements on single-crystal VO$_2$ excited by femtosecond near-IR laser pulses show transitional structural states that exist up to hundreds of picoseconds [18]. However, no evidence of structural intermediates is known for the adiabatic, thermally induced phase transition, although Qazilbash *et al*[5] have reported an intermediate electronic state characterized by strongly correlated metallic nanopuddles with properties distinct from those of the high-temperature tetragonal metal. There have also been reports of intermediate structural states of monoclinic M$_2$ and triclinic T phases [19-21] in doped or uniaxially stressed VO$_2$ indicating the existence of other local energy minima. The analysis we use in this paper for measuring, comparing and contrasting the SPT and IMT provide the first opportunity to look for intermediate structural states and correlate them to specific stages in the electronic phase transition.

We employed high-temperature x-ray diffraction (HTXRD) to follow the SPT and change in near-IR transmission as a signature of the evolution of the electronic phase transition through the strong correlation region on epitaxial thin films of VO$_2$ (see

*Supplementary Information* for details). The hysteretic responses of the IMT and SPT exhibit significantly different temperature evolutions. The dielectric function of the $VO_2$ calculated as a function of temperature from effective-medium theory using the measured volume fractions from HTXRD is also not consistent with the optical signature of the electronic transition. This supports the existence of an intermediate electronic state. These results are observed for two types of relaxed epitaxial films -- continuous epitaxial $VO_2$ (CE $VO_2$) and noncontiguous epitaxial $VO_2$ (NCE $VO_2$) films grown on c-cut sapphire having quite distinct morphologies suggesting that the differences between the structural and electronic phase transitions is a general property of epitaxial $VO_2$ thin films.

The CE and NCE $VO_2$ films of approximately 90nm thickness on c-cut (0001) sapphire substrates were grown respectively at room temperature followed by annealing, and at high temperature of 500°C following the two protocols described in *Supplementary Information*. The morphological differences in the CE and NCE films are evident from the comparison of the two SEMs respectively in figures 1 (a) and (d) and TEMs in figures (b) and (e). The continuous film grown at room temperature with subsequent annealing has a smoother surface than the discrete, pyramid-like structured NCE film grown at high temperature. But both films are epitaxial (that is have a definite orientational relationship to the substrate), as evident from the reciprocal space maps shown in figures (c) and (f) and further substantiated by figures S1 and S2.

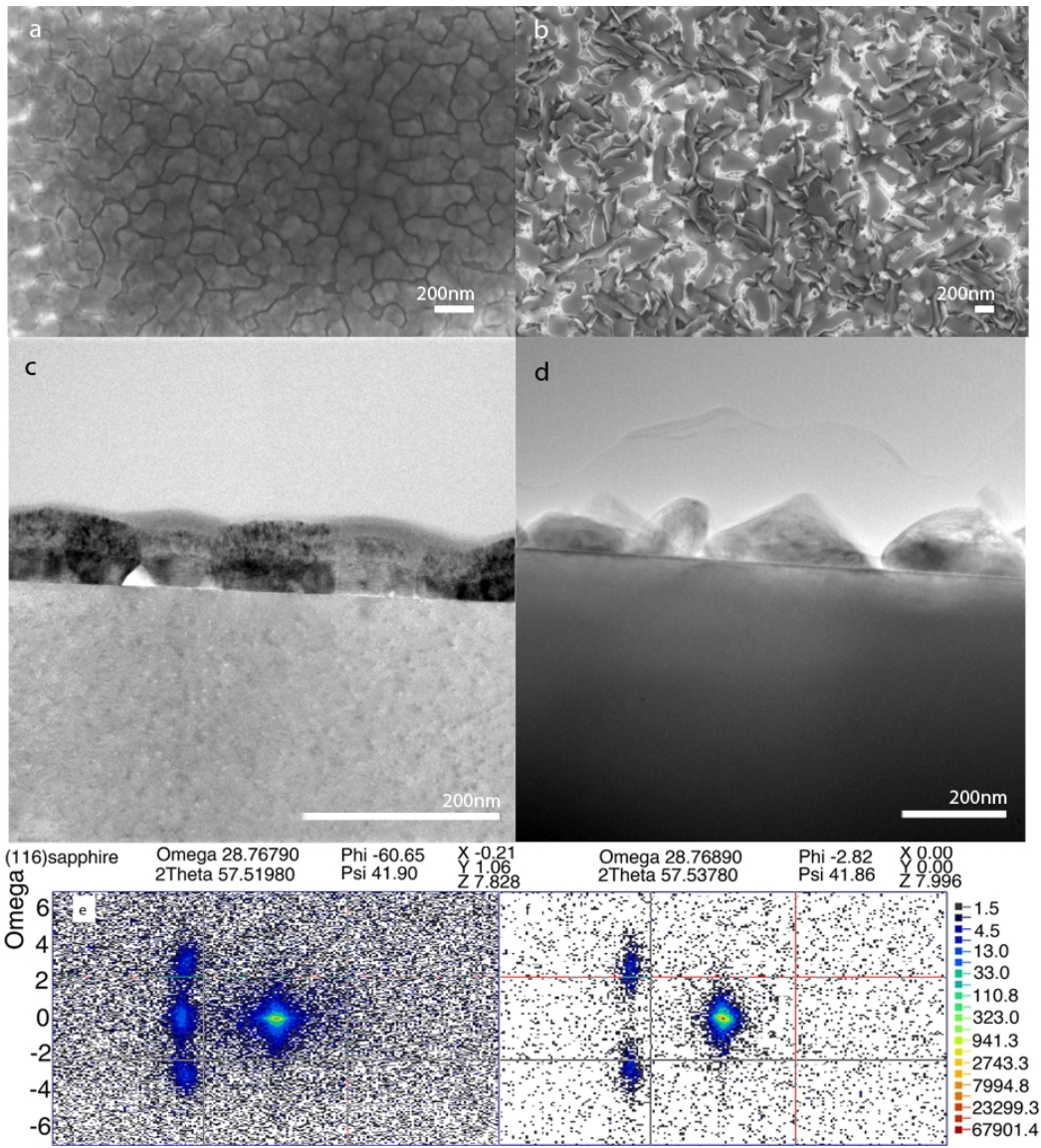

**Figure 1:** (a), (d) Top-view SEM, (b), (e) cross-sectional TEM and (c), (f) X-ray diffraction reciprocal space map of typical CE-VO$_2$ and NCE-VO$_2$ films respectively.

During the transition, we tracked the evolution of the monoclinic (040) peak for VO$_2$, located at 86.05° on the 2θ–axis for the CE VO$_2$ film; the corresponding tetragonal peak is located at 85.73°. Figure 2(a) shows the isoline 3D view of the X-ray spectra showing how the selected $VO_2$ peak shifts during the entire monoclinic-tetragonal-monoclinic transition. The superposition of the $2\theta$–spectra for the five temperatures over which almost the entire transition is completed is shown in figure 2(b). Starting at 69°C, the area under the monoclinic peak decreases as the tetragonal fraction in the film grows; evidently both phases coexist in almost equal fractions at 71°C, and the film is almost entirely tetragonal at 73°C. The slight shift in peak position during the transition can be attributed to the thermal expansion of the sample holder, substrate and the $VO_2$ lattice. Figure 2(c) shows the monoclinic fraction in the sample at each temperature, calculated from the XRD peak areas as a function of temperature (see *Supplementary Information* for details). White light transmission measurements of the MIT [Figure 2(d)] in this film show narrow, high-contrast hysteresis similar to those of bulk $VO_2$. This is the electronic signature of the phase transition.

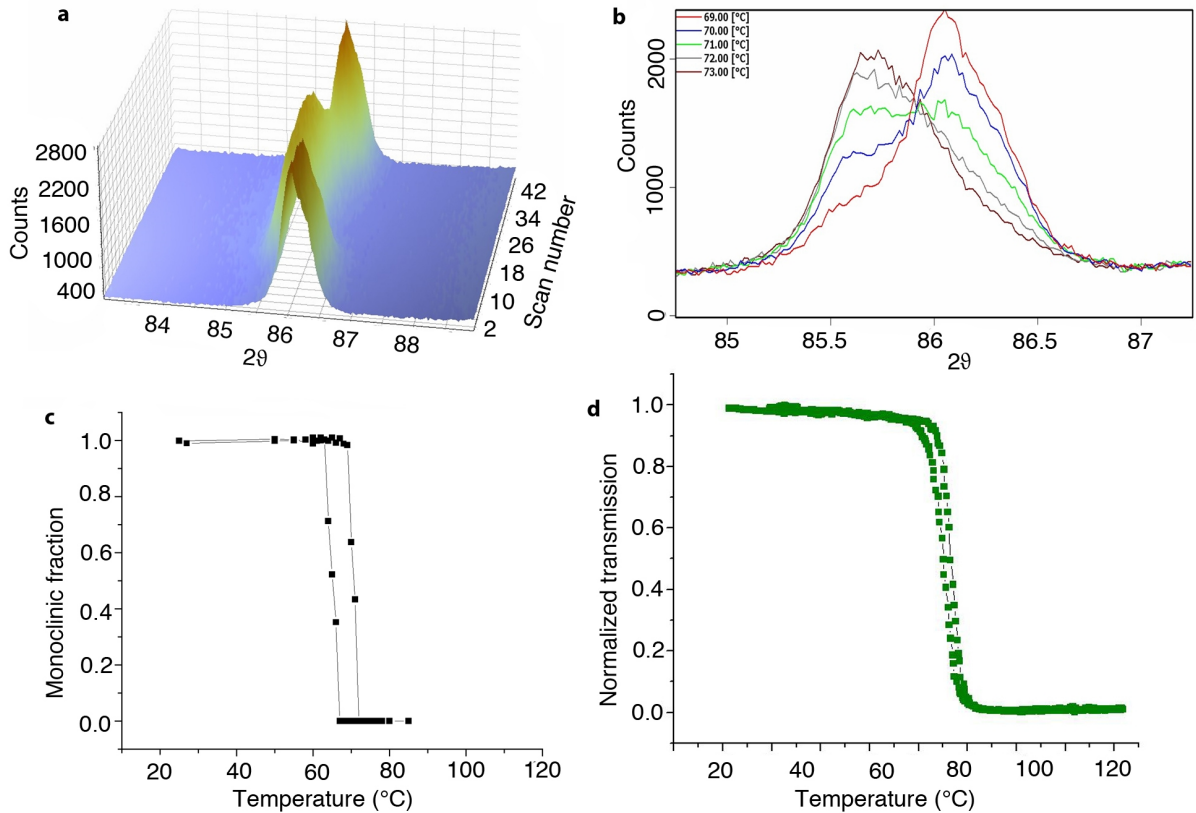

**Figure 2:** HTXRD measurements on CE VO$_2$ on c-sapphire: (a) Evolution of the reversible structural phase transition (monoclinic-tetragonal-monoclinic) as a function of temperature, (b) evolution of the monoclinic and tetragonal peaks between 69°C and 73°C, and (c) hysteresis in the monoclinic fraction as a function of temperature and (d) hysteresis in the white light optical transmission as a function of temperature measured on the same sample.

Figures 3(a) – (c) illustrate the corresponding observations for NCE-VO$_2$ films on c-cut sapphire. Here the peak at 2θ = 39.97 degrees, the (020) monoclinic peak of VO$_2$ is tracked. As the SPT starts, the monoclinic peak decreases but no tetragonal peak appears

[figures 3(a) and (b)] signaling conversion to the rutile state. Because powder diffraction XRD is designed to acquire signals from randomly oriented crystallites, we attribute this absence to the high degree of orientation characteristic of epitaxial film, for which a slight ω-offset of the tetragonal planes might cause the signal to disappear. Figure 3(d) is the experimental transmission hysteresis for the same film. For a discussion and comparison of the features in all the observed hystereses refer to the *Supplementary Information*.

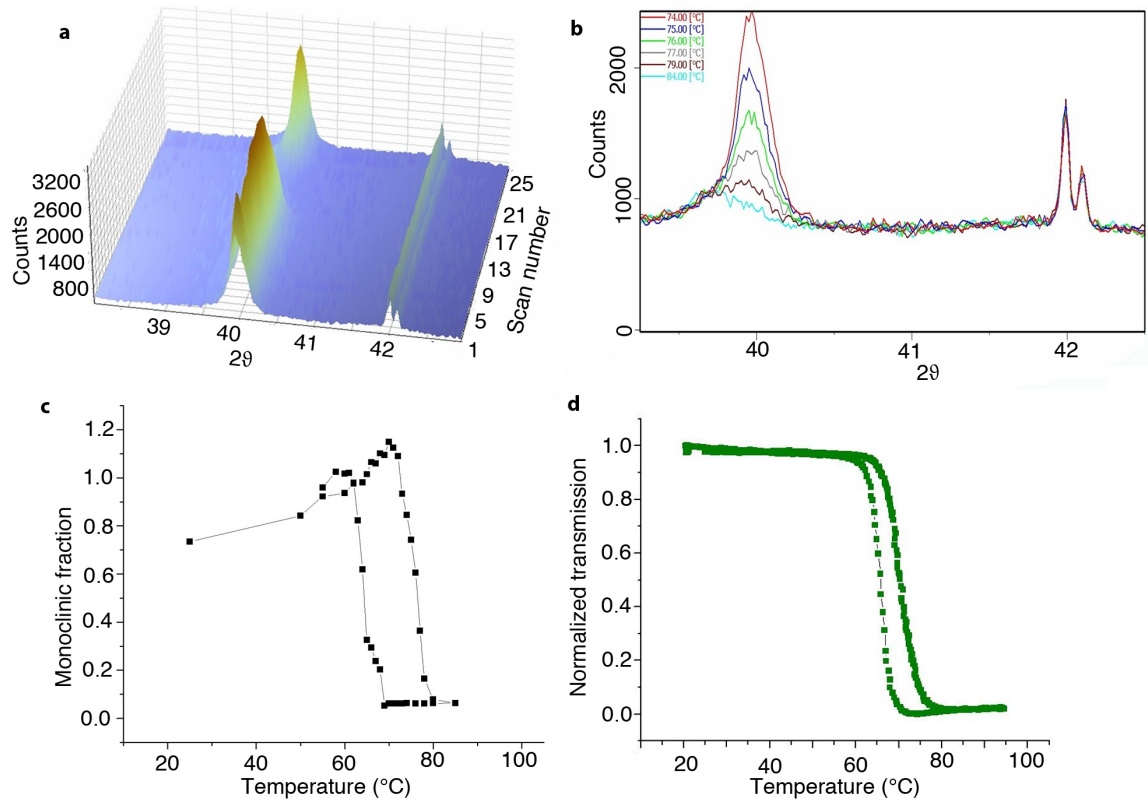

**Figure 3:** HTXRD measurements on NCE $VO_2$ on c-sapphire: (a) Evolution of the structural phase transition (monoclinic peak vanishing) as a function of temperature, (b) Evolution of the monoclinic peak between 74°C and 84°C, and (c) Hysteresis of the monoclinic phase as a function of temperature and (d) optical transmission hysteresis as a function of temperature taken on the same sample. In figures (a) and (b) the sapphire peak about 2θ ~ 41.67° remains unchanged throughout the VO$_2$ transition cycle, serving as the witness peak for the entire HTXRD measurement.

The phase transition in semiconducting VO$_2$ is initiated at nucleation sites from which metallic domains then evolve up to and beyond the transition temperature. In the strong correlation region near the transition temperature, semiconducting and metallic domains coexist, and this spatial inhomogeneity strongly influences the effective dielectric properties of the film. We use Bruggeman effective medium theory (EMT)[22] to model the optical properties of VO$_2$ thin films [23] [24] rather than Maxwell Garnett EMT [25] [26], since the latter assumes a small volume fraction of metallic inclusions. The effective dielectric constant in the Bruggeman theory for ellipsoidal inclusions is given by:

$$f \frac{\tilde{\varepsilon}_m(\lambda) - \tilde{\varepsilon}_{eff}(\lambda)}{\tilde{\varepsilon}_m(\lambda) + \frac{1-q}{q}\tilde{\varepsilon}_{eff}(\lambda)} + (1-f)\frac{\tilde{\varepsilon}_i(\lambda) - \tilde{\varepsilon}_{eff}(\lambda)}{\tilde{\varepsilon}_i(\lambda) + \frac{1-q}{q}\tilde{\varepsilon}_{eff}(\lambda)} = 0 \quad (1)$$

so that

$$\varepsilon_{eff} = \frac{\tilde{\varepsilon}_m(f-q) - \tilde{\varepsilon}_i(f+q-1) + \sqrt{[\tilde{\varepsilon}_i(f+q-1) - \tilde{\varepsilon}_m(f-q)]^2 + 4(1-q)q\tilde{\varepsilon}_m\tilde{\varepsilon}_i}}{2(1-q)} \quad (2)$$

where $\tilde{\varepsilon}_i$ and $\tilde{\varepsilon}_m$ are the wavelength-dependent dielectric constants in the insulating and metallic states of VO$_2$ respectively [27], $f$ is the metallic volume fraction obtained from the

XRD peak areas, and *q* is a shape-dependent depolarization factor for the inclusions (see *Supplementary Information* and figure S4 for details).

Figure 4 shows the real (*n*) and imaginary (*k*) parts of the refractive index of both NCE $VO_2$ and CE $VO_2$ films calculated using equation (2), and the modeled transmission using equation (3). The transmission calculated from the Bruggeman model is,

$$T_B = (1-R)e^{-\alpha} = (1-R)e^{-4\pi k z/\lambda} \qquad (3)$$

where R is the reflection coefficient (calculated from n and k), $\alpha$ is the absorption coefficient, both being functions of wavelength $\lambda$, and z is the thickness of the film, set to 90 nm in our calculations. Superimposed on the calculated transmission are the measured data from figure 2(d) and 3(d), respectively to allow comparison of the hystereses for the MIT (measured optically) and the SPT calculated from the Bruggeman analysis.

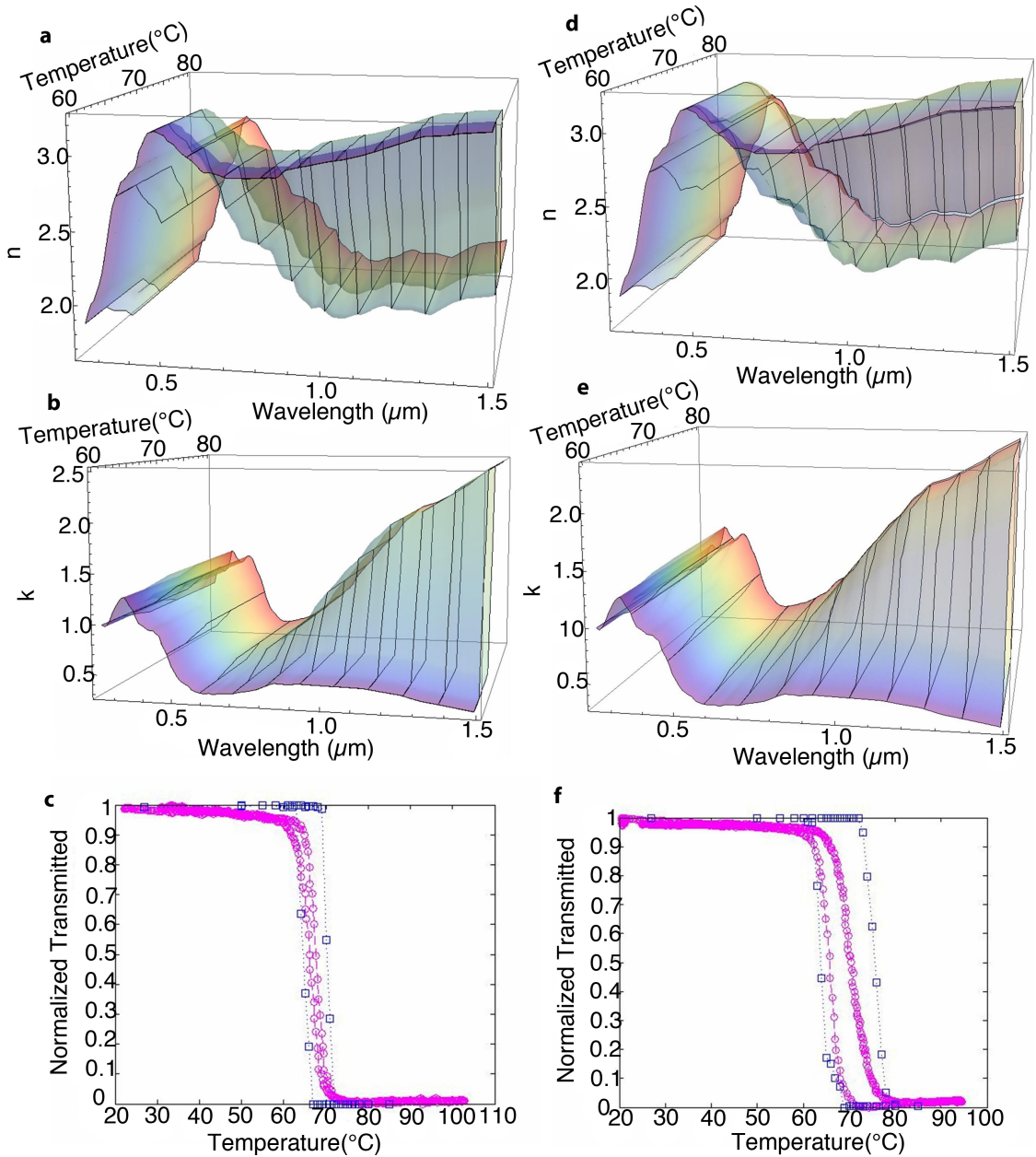

**Figure 4:** Top four panels: Three-dimensional plots of (a) real (*n*) and (b) imaginary (*k*) part of the effective refractive index in CE $VO_2$ and same in (d) and (e) respectively for NCE $VO_2$ extracted from HTXRD data by means of Bruggeman EMT analysis, as a function of temperature. The hysteresis loops are visible as one moves toward the infrared. Bottom two panels: Superposition of the transmission hystereses, modeled (blue) and experimental (magenta) for (c) NCE and (f) CE $VO_2$.

Comparison of the measured hystereses [figures 2(c),(d) and 3(c),(d)] for both CE and NCE $VO_2$ films shows that the electronic (Mott) transition requires less thermal energy than the structural phase transition. The dielectric functions extracted from the mean-field calculation of the monoclinic-to-tetragonal ratio are consistent with this picture [figure 4 (e) and (f)]. Thus, an intermediate metallic state with its higher reflectivity appears before the signature of the stable rutile structure, as inferred in [1] but coming from a complementary measurement and modeling of the SPT.

Now we turn our attention to search for a structural correlate of this intermediate electronic state. Since the known structural intermediates of $M_2$ and T phases are insulating phases, it might be possible that there are still unknown metallic crystal structures. The search for such intermediate structures would be of general significance for the physics of phase transitions, just as the discovery of the insulating $M_2$ phase helped to establish the Mott character of $VO_2$ phase transitions. It is also possible that there is no intermediate crystalline phase, and that the monoclinic structure might remain pinned until the electrons gain enough energy from fluctuations to hop from one site to another and drive the lattice rearrangement. Irrespective of the existence of intermediate crystalline phase, these results are consistent with the concept of dual correlation lengths in $VO_2$[28], where the delocalization of the electrons in the correlated metallic state corresponds to the longer of the two.

For a transient intermediate structure to exist, its lattice constants should be close to that of the two stable ones. Hence the value of 2θ at which we expect to see the signature peaks of this intermediate structure should be in our observation range. But, in fact we observe no such new peak; moreover, the sum of the fractional areas under these two peaks is constant and equal to the monoclinic peak area prior to initiation of the structural phase transition. Hence, within the resolution limit of signal-to-noise ratio and the spatial resolution of the XRD, we observe no macroscopic intermediate lattice structure. However, further investigations with nanoscale resolution are necessary to see whether the strongly-correlated metallic phase described in ref.3 has a corresponding intermediate crystal structure.

Our experimental findings demonstrate that the evolution of the SPT, when modeled by an appropriate effective-medium theory to give information on the evolving dielectric function can reveal both correlations and differences between the electronic and structural signatures of solid-solid phase transitions. For example, doping alters the electronic signature of the metal-insulator transition in $VO_2$[29] and measurements like those described here can show whether or not these changes in electronic response have structural correlates. These results also contribute to deeper understanding of phase transitions found in other Magneli phase of Ti-O and V-O[30] systems. Indeed, as interest grows in understanding strongly correlated materials at the micro- and nano-scales, clarifying the relationship of evolving structural and electronic features of the phase transitions is imperative.